\documentclass[10pt,twocolumn]{article}
\usepackage{latex8}
\usepackage{balance}
\usepackage{times}

\font\tt=cmtt10 at 12pt
\ifx\pdfoutput\undefined
\usepackage{graphicx}
\usepackage[dvips]{epsfig}
\else
\usepackage[pdftex]{graphicx}
\fi
\usepackage{subfigure}
\usepackage{amsmath}
\usepackage{array}
\usepackage{latexsym}
{\bf}{\rm}
\begin{document}
%
\pagestyle{empty}
\title{Reflective and Refractive Variables: A Model for Effective and Maintainable
Adaptive-and-Dependable Software}

\author{Vincenzo De Florio, Chris Blondia \vspace*{3pt} \\
  University of Antwerp\\
  Department of Mathematics and Computer Science\\
  Performance Analysis of Telecommunication Systems group\\
  Middelheimlaan 1, 2020 Antwerp, Belgium \vspace*{3pt} \\
  Interdisciplinary institute for BroadBand Technology\\
  Gaston Crommenlaan 8, 9050 Ghent-Ledeberg, Belgium}

\maketitle

%
\begin{abstract}
We propose a simple and effective tool for the expression of tasks such as cross-layer
optimization strategies or sensors-related applications. The approach is based on what we refer to as
``reflective and refractive variables''. Both types of variables are associated with external entities,
e.g. sensors or actuators. A reflective variable is a volatile variable, that is, a variable that might
be concurrently modified by multiple threads. A library of threads is made available, each of which
interfaces a set of sensors and continuously update the value of a corresponding set of sensors. One
such thread is ``cpu'', which exports the current level of usage of the local CPU as an integer between 0
and 100. This integer is reflected into the integer reflective variable cpu. A refractive variable is a
reflective variable that can be modified. Each modification is caught and interpreted as a request to
change the value of an actuator. For instance, setting variable ``tcp\_sendrate'' would request a
cross-layer adjustment to the thread interfacing the local TCP layer entity. This allows express in an
easy way complex operations in the application layer of any programming language, e.g. plain old C. We
describe our translator and the work we are carrying out within PATS to build simple and powerful
libraries of scripts based on reflective and refractive variables, including robotics applications and
RFID tags processing.
\end{abstract}


\Section{Introduction}
\thispagestyle{empty}
As well known, a number of problems require solutions that involve the whole of the system layers, from
the bare machine up to the application. Problems of this type include e.g. fault-tolerance, cross-layer
signaling, or adaptability~\cite{SaRC84}. We can observe that wherever there is a need for flexibility,
performance, quality trade-offs, or security and co-operation, there exists a need to monitor and adjust
parameters across the whole of the system layers. Much more than this, there exist a need to do so in an
as much as possible simple way, from both an architectural and the user points of view. This means that
the architectural strategy must be simple and that the way to express the solutions must be
straightforward.  To date, several clever architectural strategies to solve those problems exist. Just
to name a few, the energy-performance manager of IMEC~\cite{BLE03} or the network-status of 
Mobiman~\cite{CMTS04} provide
interesting architectures to reach effective cross-layer optimization. In both the mentioned approaches,
though, no solution is envisaged to the problem of the optimal expression of cross-layered adaptations.  
For instance, both the above mentioned approaches require ad hoc versions of the protocol layers,
versions that explicitly make use of the network status. Each layer, to be compliant to these models,
must endorse logics to take actions making use of the information kept in a network status
database. This requires the design of ad hoc software. In such software the two concerns -- the
functional one, i.e. the layer function, and the non-functional one, for cross layer adaptation -- are
mixed and intertwined. A possible solution currently being investigated by other researchers is aspect
oriented computing~\cite{KLM97,Bon04}, which requires the use of custom programming languages and complex
tools.  We propose a simpler, language independent solution that we call reflective and refractive
variables (in short, RR vars). In the following we describe our approach in Sect.~2 and we show how we
implemented it in Sect.~3. An analysis of current and possible uses of RR vars in fields such as
robotics, sensor networks, and RFID applications, is presented in Sect.~4. Section 5 concludes this work
with a description of our future plans involving RR vars.

\Section{Reflective, Refractive and Redundant Variables}
The idea behind RR vars is to use memory access as an abstraction to perform concealed tasks. RR vars
are volatile variables whose identifier links them with an external device, such as a sensor, or an
RFID, or an actuator. In reflective variables, memory cells get asynchronously updated by service
threads that interface those external devices. We use the well-known concept of reflection because those
variables ``reflect'' the values measured by those devices. In refractive variables, on the contrary, write
requests trigger a request to update an external parameter, such as the data rate of the local TCP
protocol entity or the amount of redundancy to be used in transmissions. We use to say that write
accesses ``refract'' (that is, get redirected~\cite{TST}) onto corresponding external devices.

\begin{figure}
\includegraphics[width=0.5\textwidth]{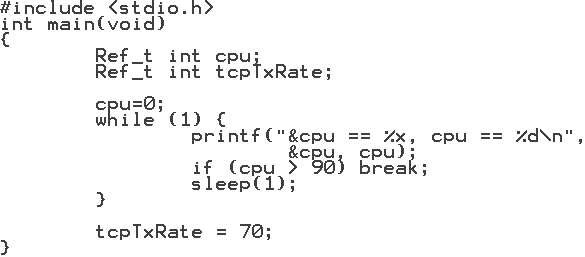}
\caption{A simple example of the use of RR vars.}
\end{figure}

The RR var model does not require any special language: Figure 1 is an example in the C language. 
The portrayed
program declares two variables: ``cpu'', a reflective integer, which reports the
current level of usage of the local CPU as an integer number between 0 and 100,
and ``tcpTxRate'', a reflective {\em and refractive\/} integer, which reports {\em and sets\/}
the send rate parameter of the TCP layer. The code periodically queries
the CPU usage and, when that reaches a value greater than 90\%, it requests to
change the TCP send rate. Note that the only non standard C construct is
attribute ``Ref\_t'', which specifies that a corresponding declaration is
reflective or refractive or both. Through a translation process, discussed in
Sect.~3, this code is instrumented so as to include the logics required to
interface the cpu and the TCP external devices. Figure 3 shows this simple code
in action on our development platform---a Pentium-M laptop running Windows
XP and the Cygwin tools.

\begin{figure*}[t]
\centerline{\includegraphics[width=0.75\textwidth]{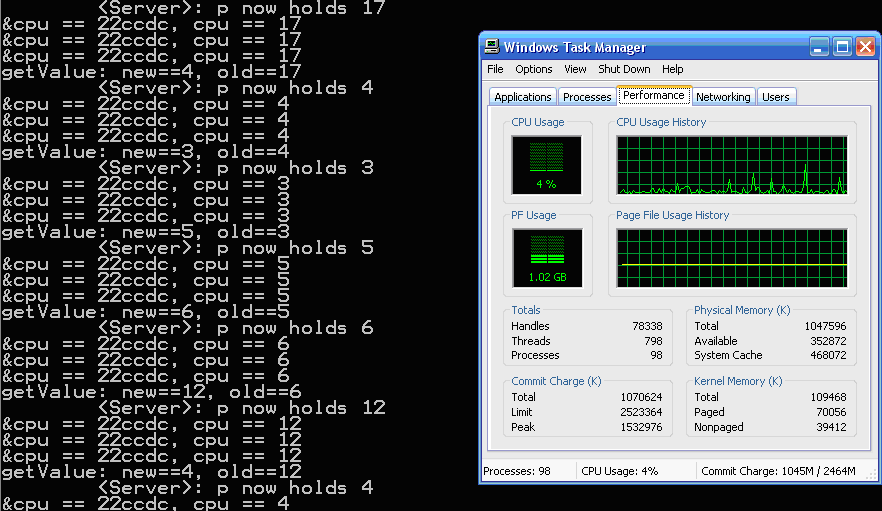}}
\caption{An excerpt from the execution of the code in Fig.~2.}
\end{figure*}

We observe that through the RR var model the design complexity is partitioned into two well defined and
separated components: the code to interface external devices is specified ``elsewhere'' (Sect.~3 describes
where and how) while the functional code is specified in a familiar way, in this case as a C code
reading and writing integer variables.

The result is a structured model to express tasks such as cross-layered optimization, adaptive or
fault-tolerant computing in an elegant, non intrusive, and cost-effective way. Such model is
characterized by strong separation of design concerns, for the functional strategies are not to be
specified aside with the layer functions; only instrumentation is required, and this can be done once
and for all. This prevents spaghetti-like coding for both the functional and the non-functional aspects,
and translates in enhanced maintainability and enhanced efficiency.

The RR var model provides the designer also with another attribute: a variable, be it an RR var or a
``common'' one, can be tagged as being ``\emph{redundant}''. Redundant variables are variables whose contents
get replicated several times so as to protect them from memory faults. Writing to a redundant variable
means writing to a number of replicas, either located strategically\footnote{Strategically means here that
the redundant cells are allocated in such a way as to tolerate possible burst errors, affecting
contiguous memory cells.} on the same processing node or on remote nodes---when available and the extra
overhead be allowed. Reading from a redundant variable actually translates in reading from each of its
cells and performing majority voting. The result of this process is monitored by a special device, that
we call Redundance. Redundance measures the amount of votes that differ from the majority vote, and uses
this as a measure of the disturbance in the surrounding environment. Under normal situation, Redundance
triplicates the memory cells of redundant variables. This corresponds to tolerating up to one memory
fault in the cells associated to a redundant variable. Under more critical situations, the amount of
redundancy should change. This is what actually happens: the component that manages redundant variables
declares the integer reflective variable ``ref\_t int redundance''. The latter is set asynchronously by
the Redundance device, which adjusts the corresponding memory cells\footnote{Not surprisingly enough,
variable ``redundance'' is indeed\ldots{} redundant.} with a number representing the ideal degree of
redundancy with respect to the current degree of disturbances.

The RR var model does not support only cross-layer optimization---in general, it provides an
application-layer construct to manage feedback loops.

\begin{figure*}
\centerline{\includegraphics[width=0.75\textwidth]{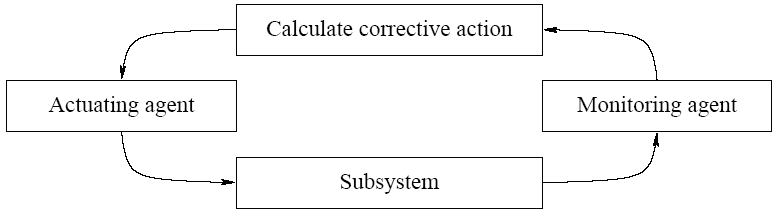}}
\caption{General structure of feedback loops (picture from~\cite{PVR06}).}
\end{figure*}

Feedback loops (see Fig.~3)---a well known concept from system theory are ideal forms to shape our systems
so as to be adaptive-and-dependable~\cite{PVR06}. Such property is an important pre-requisite for the
welfare of our computer-dominated societies and economies: in the cited paper Van Roy explains their
relevance to future software design. RR vars provide a straightforward syntactical structure and
software architecture for the expression of feedback loops. We use this structure, e.g., to implement
redundant variables. The main advantage in this case is that, instead of taking a design decision once
and for all, we let a system parameter change as needed, zeroing in on the optimum. The use of RR vars
simplifies the design of our solution, which also enhances maintainability. But probably the most
important consequence is that our solution does not assume a fixed, immutable fault model, but lets it
change with the actual faults being experienced.

\begin{figure*}
\centerline{\includegraphics[width=0.75\textwidth]{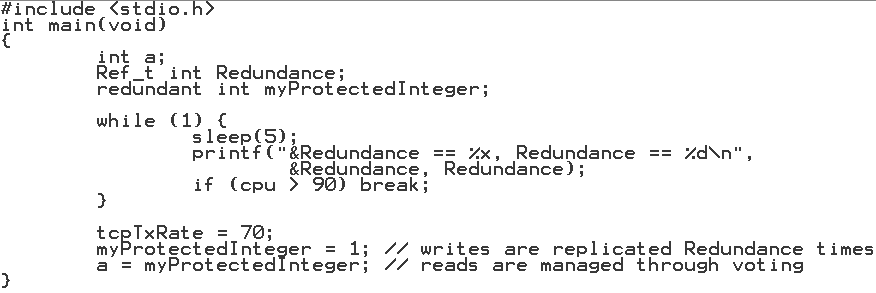}}
\caption{Redundant variables.}
\end{figure*}

Figure 4 shows how simple it is using a redundant variable: no syntactic differences can be noticed. The required logic is ``hidden'' in the translation process.

\Section{Implementation}
The core of the RR vars architecture is a parser that translates the input source code into two source files, one with an augmented version of the original code and one server-side to monitor and drive the external devices. To explain this process we consider Fig.~5, an excerpt from the translation of the code in Fig.~4.
Let us
review the resulting code in more detail (please note that item x in the following list refer to lines
tagged as ``// {\em x}'' in the code):

\begin{enumerate}
 \item First the translator removes
     the occurrences of attributes ``ref\_t'' and ``redundant''.
 \item Then it performs a few calls to
     function ``aopen''. This is to open the associative arrays ``reflex'' and ``rtype''.
     As well known, an associative array generalizes the concept of array so as
     to allow addressing items by non-integer indexes. The arguments to ``aopen''
     are functions similar to ``strcmp'', from the C standard library, which are
     used to compare index objects. The idea is that these data structures
     create links between the name of variables and some useful information (see
     below).
 \item There follow a number of
     ``awrites'', i.e., we create associations between variables identifiers and
     two numbers: the corresponding variables' address and an internal code
     representing its type and attributes.
 \item Then ``Server'', the thread
     responsible to interface the external devices, is spawned.
 \item Besides a write access into
     refractive variable tcpTxRate, the translator places a call to function
     ``CalltcpTxRate''. In general, after a call to refractive variable {\em v},
     the call ``Call{\em v}(\&{\em v})'' is produced.
 \item Similarly, a write access to
     redundant variable {\em w}, of type {\em t}, is followed by a call to
     ``RedundantAssign\_{\em t}(\&{\em w})''.
 \item Finally, reading from redundant
     variable {\em w}, of type {\em t}, is translated into a call to
     function ``RedundantRead\_{\em t}(\&{\em w})''.
\end{enumerate}

\begin{figure*}
\centerline{\includegraphics[width=0.739\textwidth]{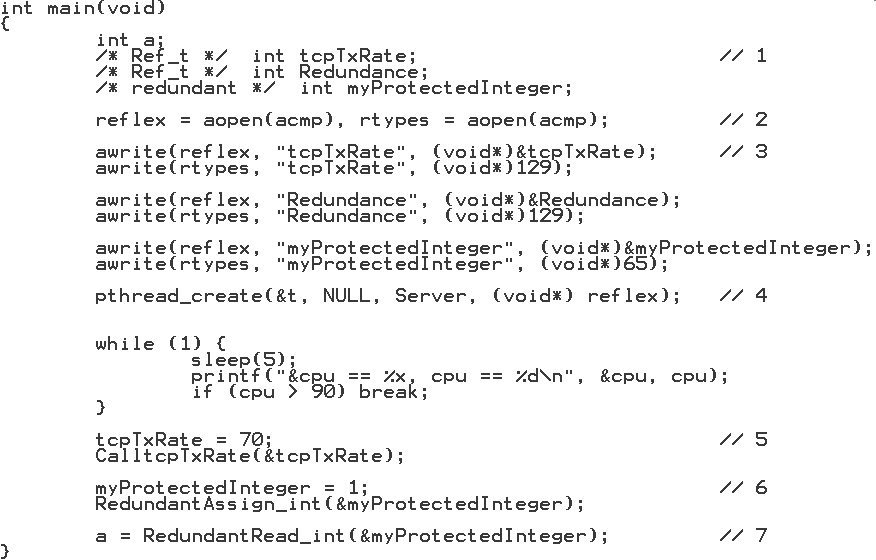}}
\caption{Abridged version of the main function of the translated code.}
\end{figure*}

It is the
responsibility of the designer to make sure that proper code for functions
``Call{\em v}(\&{\em v})'' is produced. Functions ``RedundantAssign\_{\em t}(\&{\em w})'' and
``RedundantRead\_{\em t}(\&w)'' are automatically generated through
a template-like approach---the former performs a redundant write, the latter a
redundant read plus majority voting. For voting, an approach similar to that 
in~\cite{DeDL98e} is followed. Associative arrays are managed through class
ASSOC~\cite{Dev27}.

As already
mentioned, the ``Server'' thread is the code responsible to monitor and interface
the external devices. Its algorithm is quite simple (see Fig.~6): the code
continuously waits for a sensor update (lines tagged with ``// 1''), then
retrieves the address and type of the corresponding reflective variable (in 
``// 2'') and finally updates that variable (``// 3'').

\begin{figure*}
\centerline{\includegraphics[width=0.739\textwidth]{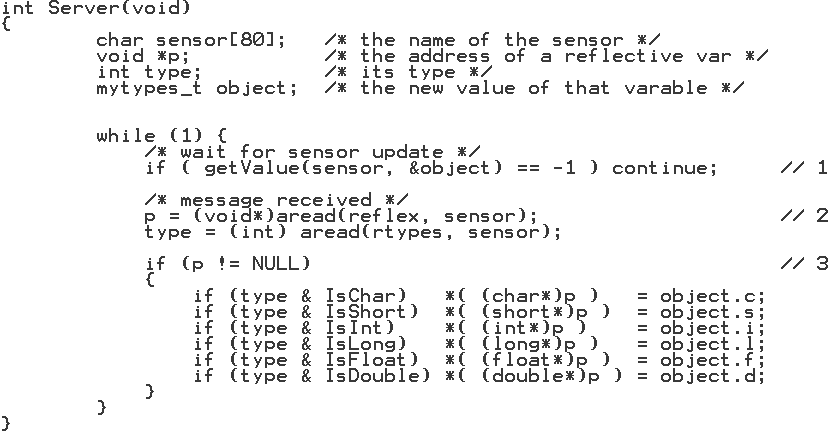}}
\caption{The Server code.}
\end{figure*}

The
complexity to interface external devices is charged to function ``getValue'', 
an excerpt of which is shown
in Fig.~7. The core of ``getValue'' is function ``cpu'',
which returns the amount of CPU currently being used.

\begin{figure*}
\centerline{\includegraphics[width=0.7\textwidth]{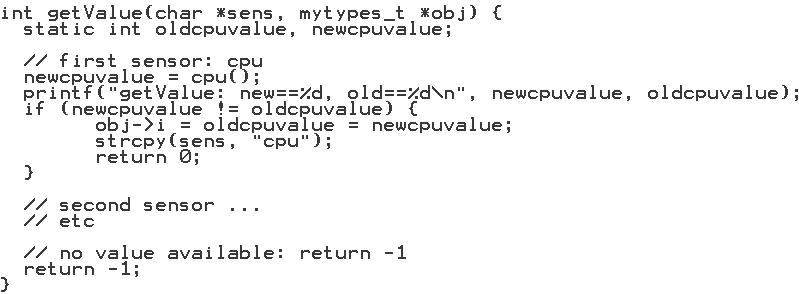}}
\caption{Function getValue interfaces all the external devices that are connected to RR vars.}
\end{figure*}

\Section{Problem Solving with RR Vars}
We are in
the process of making use of RR vars in several real-life applications---we
plan to report on these use cases in future papers. In the meanwhile we report
herein on possible contexts where RR vars could provide effective and low-cost
solutions.

\SubSection{Concurrency}
As cleverly explained e.g.
by Gates in~\cite{Gat07}, a well known challenge in robotics is {\em concurrency}, defined in the
cited paper as ``how to simultaneously
handle all the data coming in from multiple sensors and send the appropriate
commands to the robot's motors''. The conventional approach, i.e., making use of
a long loop that first reads all the data from the sensors, then processes the input
and finally controls the robot is not adequate enough. Because of this, the
robot control could be using stale values, which could bring to disastrous
consequences. As Gates mentions in the cited paper, this is a scenario that
applies not only to robotics but also to all those fields such as distributed
and parallel computing where data and control often need to be effectively
orchestrated under strict real-time constraints. ``To fully exploit the power of
processors working in parallel, the new software must deal with the problem of
concurrency'', Gate says. We believe an approach like RR vars can be an effective syntactic
structure for that: a control loop using reflective variables, for instance, would
not need to specify a reading order for the input variables, which are updated asynchronously, as new values 
need to replace old ones.

\SubSection{Localizing Hidden Assets}
We are
currently extending our translator so as to allow writing programs such as the
one in Fig.~8.

\begin{figure}[t]
\centerline{\includegraphics[width=0.35\textwidth]{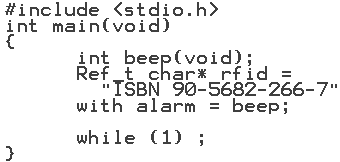}}
\caption{RR var to localize objects with RFID tags on them.}
\end{figure}

At first
sight the program may sound meaningless, as it only declares a function and an
RR var, ``rfid'', and does not seem to perform any useful action. ``Behind the
lines''---a nice feature offered by translators---what happens is that
surrounding RFID tags reflect their content onto reflective variable ``rfid''. Data
stored into that variable is compared with the initialization value (in this
case, an ISBN number). In case of a match, function ``beep'' is called.

Now
imagine running this code onto your PDA while walking through the lanes of a
large library such as the Vatican Library in search for a ``lost'' or misplaced
book. When in reach of the searched item, the PDA starts beeping\footnote{The tomes of the Vatican library have been recently equipped with RFID tags.}.
Or imagine that, thanks to international regulations, all ``companies'' building
antipersonnel mines be obliged by law to embed RFID tags into their ``products''.
When activated, these tags and a program as simple as the one in Fig.~8 could
easily prevent dreadful events that continuously devastate the lives of too many a human being.

\Section{Conclusions}\label{s:end}

We
introduced a translation system that allows making use of reflection in a
standard programming language such as C. The same translator supports
``refraction'', that is the control of external devices through simple memory
write accesses. These two features are used to realize redundant data
structures. As well known, redundancy is a key property in
fault-tolerance. The Shannon teorem teaches us that
through any unreliable channel it is possible to send data reliably by using a
proper degree of redundancy. This famous result can be read out in a different
way: for each degree of unreliability, there is a minimum level of redundancy
that can be used to tolerate any fault. Our approach uses RR vars to attune the
degree of redundancy required to ensure data integrity to the actual faults
being experienced by the system. This provides an example of adaptive
fault-tolerant software.

RR vars
can be used to express problems in cross-layer optimization, but also in
contexts where concurrency calls for expressive software structures, e.g.
robotics. Localization problems could also be solved through a very simple scheme.
Other fields where we are exercizing our tool include personalized healthcare~\cite{DB07e}
and global adaptation frameworks to enhance the quality of experience of mobile services~\cite{DB07f}.
Within PATS we are now further improving our model and tools and designing a
few simple and powerful libraries of scripts based on reflective and refractive
variables.

\section*{Acknowledgement.} The reported work took place in the framework of our participation
to IBBT~\cite{ibbt} project ``End-to-end Quality of Experience''~\cite{E2EQoE} and
IST-NMP2-016880 project ARFLEX (``Adaptive Robots for FLEXible manufacturing systems'')~\cite{Arflex}.
In particular it is our pleasure to acknowledge the ever valuable suggestions of Prof. Lucio Businaro of
EICAS.

\balance

\bibliographystyle{latex8}

\end{document}